\documentclass{elsart}
\usepackage{epsfig}
\newlength{\wordlength}

\newlength{\onewordlength}

    \newcommand{\ba}{\begin{eqnarray}}
    \newcommand{\ea}{\end{eqnarray}}
    \newcommand{\be}{\begin{equation}}
    \newcommand{\ee}{\end{equation}}

    \newcommand{\AmS}{{\protect\the\textfont2%
  A\kern-.1667em\lower.5ex\hbox{M}\kern-.125emS}}
  \newcommand{\oneudot}[2]{\settowidth{\wordlength}{#1}%
    \setlength{\onewordlength}{\wordlength}%
    \mathop{\mbox{#1}} \limits_{#2}}
  \newcommand{\tcirc}[1]{{\stackrel{\circ}{#1}}}

\newcommand{\bzero}{{\bf 0}}

\newcommand{\bn}{{\bf n}}
\newcommand{\tn}{{\tilde{\bn}}}
\newcommand{\hn}{{\hat{\bn}}}
\newcommand {\bk} {{\mathbf k}}
\newcommand {\bfr} {{\mathbf r}}
\newcommand {\bp} {{\mathbf p}}

\newcommand{\bx}{{\bf x}}

\newcommand{\calH}{{\mathcal H}}
\newcommand{\calK}{{\mathcal K}}

\newcommand{\calO}{{\mathcal O}}
\newcommand{\calM}{{\mathcal M}}
\newcommand{\calZ}{{\mathcal Z}}
\newcommand{\calY}{{\mathcal Y}}
\newcommand{\calW}{{\mathcal W}}

\begin{document}
\runauthor{PKU}
\begin{frontmatter}

\title{Two Particle States in an Asymmetric Box and the Elastic Scattering Phases}
\author[PKU]{Xu Feng},
\author[PKU]{Xin Li}
\author[PKU]{and Chuan Liu}

\address[PKU]{Department of Physics\\
          Peking University\\
                  Beijing, 100871, P.~R.~China}
                \thanks{This work is supported by the National Natural
 Science Foundation (NFS) of China under grant
 No. 90103006, No. 10235040 and supported by the
 Trans-century fund from Chinese
 Ministry of Education.}

\begin{abstract}
The exact two-particle energy eigenstates in a generic asymmetric
rectangular box with periodic boundary conditions in all three
directions are studied. Their relation with the elastic scattering
phases of the two particles in the continuum are obtained for both
$D_4$ and $D_2$ symmetry. These results can be viewed as a
generalization of the corresponding formulae in a cubic box
obtained by L\"uscher before.  In particular, the $s$-wave
scattering length is related to the energy shift in the finite
box. Possible applications of these formulae are also discussed.
\end{abstract}
\end{frontmatter}


\section{Introduction}

 Scattering experiment serves as a major
 experimental tool in the study of interactions among particles.
 In these experiments, scattering cross sections are
 measured. By a partial wave analysis, one obtains the
 experimental results on particle-particle scattering
 in terms scattering phase shifts in a channel of
 definite quantum numbers.
 In the case of strong interaction, experimental results on
 hadron-hadron scattering phase shifts are available in
 the literature. On the theoretical side, Quantum Chromodynamics (QCD)
 is believed to be the underlying theory of strong interactions. However, due
 to its non-perturbative nature, low-energy hadron-hadron
 scattering should be studied with a non-perturbative
 method. Lattice QCD provides a genuine non-perturbative method which
 can tackle these problems in principle, using numerical simulations.
 In a typical lattice calculation, energy eigenvalues of two-particle states
 with definite symmetry can be obtained by measuring appropriate
 correlation functions. Therefore, it would be desirable to
 relate these energy eigenvalues which are available through
 lattice calculations to the scattering phases which are obtained
 in the scattering experiment.  This was accomplished
 in a series of papers by L\"uscher
 \cite{luscher86:finiteb,luscher90:finite,luscher91:finitea,luscher91:finiteb}
 for a cubic box topology.
 In these references, especially Ref.~\cite{luscher91:finitea},
 L\"uscher found a non-perturbative relation of
 the energy of a two-particle state in a cubic box (a
 torus) with the corresponding elastic scattering phases of the two
 particles in the continuum. This formula, now known as
 L\"uscher's formula, has been utilized in a number of
 applications, e.g. linear sigma model in the broken phase \cite{Zimmermann94},
 and also in quenched QCD~\cite{gupta93:scat,fukugita95:scat,jlqcd99:scat,%
 JLQCD02:pipi_length,chuan02:pipiI2,juge03:pipi_length,CPPACS03:pipi_phase,ishizuka03:pipi_length}.
 Due to limited numerical computational power,
 the $s$-wave scattering length, which is related to the scattering
 phase shift at vanishing relative three momentum,
 is mostly studied in hadron scattering using quenched approximation.
 CP-PACS collaboration calculated the scattering
 phases at non-zero momenta in pion-pion $s$-wave scattering in the $I=2$
 channel \cite{CPPACS03:pipi_phase} using quenched Wilson
 fermions and recently also in two flavor
 full QCD \cite{CPPACS03:pipi_phase_unquench}.

 In typical lattice QCD calculations,
 if one would like to probe for physical
 information concerning two-particle states with non-zero relative three momentum,
 large lattices have to be used which usually requires
 enormous amount of computing resource.
 One of the reasons for this difficulty is the following.
 In a cubic box, the three momenta of a single particle
 are quantized according to:
 $\bk=(2\pi/L)\bn\equiv(2\pi/L)(n_1,n_2,n_3)$,
 with $\bn\in Z^3$.
 \footnote{We use the notation $Z^3$ to stand for the set
 of three-dimensional integers. That is, $\bn\in Z^3$ means that
 $\bn=(n_1,n_2,n_3)$ with $n_1$, $n_2$
 and $n_3$ integers.}
 In order to control lattice artifacts due to these
 non-zero momentum modes, one needs to have large values of $L$.
 One disadvantage of the cubic box
 is that the energy of a free particle with lowest non-zero momentum
 is degenerate. This means that the second lowest
 energy level of the particle
 with non-vanishing momentum corresponds to $\bn=(1,1,0)$.
 If one would like to measure these states
 on the lattice, even larger values of $L$ should be used.
 One way to remedy this is to use
 a three dimensional box whose shape is not cubic.
 If we use a {\em generic} rectangular box of
 size $(\eta_1L)\times(\eta_2L)\times L$ with $\eta_1$ and
 $\eta_2$ other than unity, we
 would have three different low-lying one-particle energy with
 non-zero momenta corresponding to $\bn=(1,0,0)$, $(0,1,0)$
 and $(0,0,1)$, respectively.
 This scenario is useful in practice since it presents more available low
 momentum modes for a given lattice size,
 which is important in the study of hadron-hadron scattering
 phase shift. Similar situation also occurs in the study of
 $K$ to $\pi\pi$ matrix element (see Ref. \cite{ishizuka02:Kpipi_review} for
 a review and references therein). There, one also needs to study
 two-particle states with non-vanishing relative three-momentum.
 Again, a cubic box yields too few available low-lying
 non-vanishing momenta and large value of $L$ is needed to
 reach the physical interesting kinematic region.
 In all of these cases,
 one could try an asymmetric rectangular box with only
 one side being large while the other sides moderate.
 One only has to choose the parameter $\eta_1$ and $\eta_2$
 appropriately such that more low-lying momentum modes
 can be measure on the lattice with controllable lattice
 artifacts.

 In an asymmetric rectangular box, the original formulae
 due to L\"uscher, which give the relation between the
 energy eigenvalues of the two-particle states
 in the finite box and the continuum scattering
 phases, have to be modified accordingly.
 The purpose of this paper is to derive the equivalents of
 L\"uscher's formulae in the case of a generic
 rectangular (not necessarily cubic) box.

 We consider two-particle states in a box of size
 $(\eta_1L)\times (\eta_2L)\times L$ with periodic boundary
 conditions in all three directions. For definiteness, we take
 $\eta_1\geq 1$,$\eta_2\geq 1$, which amounts to denoting the
 the length of the smallest side of the rectangular box as $L$.
 The following derivation depends heavily on the previous
 results obtained in Ref.~\cite{luscher91:finitea}.
 We will take over similar assumptions as in Ref.~\cite{luscher91:finitea}.
 In particular, the relation between
 the energy eigenvalues and the scattering
 phases derived in the non-relativistic quantum mechanical model can be carried
 over to the case of relativistic, massive field theory under these assumptions,
 the same way as in the case of cubic box which was
 discussed in detail in Ref.~\cite{luscher91:finitea}.
 For the quantum mechanical model,
 we assume that the range of the interaction, denoted by $R$,
 of the two-particle system is such that $R<L/2$.

 The modifications which have to be implemented,
 as compared with Ref.~\cite{luscher91:finitea},
 are mainly concerned with different symmetries of the box.
 In a cubic box, the representations
 of the rotational group are decomposed into irreducible
 representations of the cubic group. In a generic asymmetric box,
 the symmetry of the system is reduced. In the case
 of $\eta_1=\eta_2\neq 1$, the basic group becomes $D_4$;
 if $\eta_1\neq \eta_2 \neq 1$, the symmetry is further reduced to $D_2$,
 modulo parity operation.
 Therefore, the final expression relating the energy eigenvalues of
 the system and the scattering phases will be different.

 This paper is organized as follows. In section~\ref{sec:SPS}, we
 discuss the singular periodic solutions to the Helmholtz
 equation. The energy eigenstates of the two-particle system
 can be expanded in terms of these solutions.
 In section~\ref{sec:symmetry}, we discuss in detail the
 symmetry of an asymmetric box. Two cases are studied:
 $\eta_1=\eta_2$ in which case the basic symmetry group
 is $D_4$ and $\eta_1\neq\eta_2$ in which case the
 symmetry group is $D_2$. The irreducible representations of the
 rotational group are decomposed into irreducible
 representations of these point groups. Energy eigenvalues
 in the $A^+_1$ sector are related to the scattering
 phases for the two cases respectively.
 In section~\ref{sec:small_q}, we discuss the low-momentum and
 large-volume limit of the general formulae obtained in
 section~\ref{sec:symmetry}. A simplified formula is
 obtained for the scattering length and numerical values
 for the coefficients of this expansion are listed.
 Finally in section~\ref{sec:conclude}, we conclude with
 some general remarks. Some details of the calculation
 are provided in the appendices for reference.

 \section{Energy eigenstates and singular periodic solutions of Helmholtz equation}
 \label{sec:SPS}

 Our notations close follow those used in Ref.~\cite{luscher91:finitea}.
 The energy eigenstates in a periodic box is intimately related to the singular
 periodic solutions of the Helmholtz equation:
 \be
 (\nabla^2+k^2)\psi(\bfr)=0 \;.
 \ee
 If the function $\psi(\bfr)$ is a solution to
 the Helmholtz equation for $\bfr\neq 0$; and it is periodic:
 $\psi(\bfr+\hat{\bn}L)=\psi(\bfr)$;
 and it is bounded by powers of $r^\Lambda$ near $r=0$;
 we call $\psi(\bfr)$ a singular periodic solution to the
 Helmholtz equation of degree $\Lambda$.

 The momentum modes in the rectangular box are quantized as:
 $\bk=(2\pi/L)\tilde{\bn}$.
 For every $\bn=(n_1,n_2,n_3)\in Z^3$,
 we introduce the notations:
 \be
 \label{eq:bn_hn_def}
 \tilde{\bn}\equiv (n_1/\eta_1,n_2/\eta_2,n_3)\;, \;\;
 \hat{\bn}\equiv (n_1\eta_1,\eta_2n_2,n_3) \;.
 \ee
 When discussing the singular periodic solutions to Helmholtz equation,
 one should differentiate two cases:
 regular values of $k$, which means that $|k|\neq (2\pi/L)|\tilde{\bn}|$ for
 any $\bn\in Z^3$ and singular values of $k$, which means that
 $|k|=(2\pi/L)|\tilde{\bn}|$ for some $\bn\in Z^3$.
 For our purpose, it suffices to study the regular values of $k$.
 In this case, the singular periodic solutions
 of Helmholtz equation can be obtained from the Green's function:
 \be
 G(\bfr;k^2)={1\over\eta_1\eta_2L^3}\sum_\bp
 {e^{i\bp\cdot\bfr}\over \bp^2-k^2}\;.
 \ee
 One can easily check that the function $G(\bfr;k^2$ is a
 singular periodic solution of Helmholtz equation with
 degree $1$. More singular periodic solutions can be obtained
 as follows. We define:
 \be
 \calY_{lm}(\bfr)\equiv r^lY_{lm}(\Omega_\bfr)\;,
 \ee
 where $\Omega_\bfr$\ represents the solid angle
 parameters $(\theta,\phi)$ of $\bfr$ in spherical coordinates;
 $Y_{lm}$ are the usual spherical harmonic functions.
 It is well-known that $\calY_{lm}(\bfr)$ consist
 of all linear independent, homogeneous functions in $(x,y,z)$ of
 degree $l$ that transform irreducibly under the rotational group.
 We then define:
 \be
 G_{lm}(\bfr;k^2)=\calY_{lm}(\nabla)G(\bfr;k^2)\;.
 \ee
 One can show that the functions $G_{lm}(\bfr;k^2)$
 form a complete, linear independent set of
 functions of singular periodic solutions of the
 Helmholtz equation with degree $l$. That is to say,
 any singular periodic solution of the Helmholtz equation
 with degree $\Lambda$ is given by
 \be
 \psi(\bfr)=\sum^\Lambda_{l=0}\sum^l_{m=-l}
 v_{lm}G_{lm}(\bfr,k^2)\;,
 \ee
 with complex coefficients $v_{lm}$.
 The functions $G_{lm}(\bfr;k^2)$ may be expanded into
 usual spherical harmonics with the result:
 \be
 \label{eq:expand_Ylm}
 G_{lm}(\bfr;k^2)={(-)^lk^{l+1}\over 4\pi}
 \left[Y_{lm}(\Omega_\bfr)n_l(kr)
 +\sum_{l'm'}\calM_{lm;l'm'}Y_{l'm'}(\Omega_\bfr)
 j_{l'}(kr)\right]\;.
 \ee
 Here, $j_l$ and $n_l$ are the usual spherical Bessel functions
 and the matrix $\calM_{lm;l'm'}$ is related to
 the {\em modified} zeta function via:
 \ba
 \label{eq:calM-zeta}
 \calM_{lm;js}&=&
 \sum_{l'm'}{(-)^si^{j-l}\calZ_{l'm'}(1,q^2;\eta_1,\eta_2)\over
 \eta_1\eta_2\pi^{3/2}q^{l'+1}}
 \sqrt{(2l+1)(2l'+1)(2j+1)}
 \nonumber \\
 &\times &
 \left(\begin{array}{ccc}
 l & l' & j \\
 0 & 0  & 0 \end{array}
 \right)
 \left(\begin{array}{ccc}
 l & l' & j \\
 m & m' & -s \end{array}
 \right)\;.
 \ea
 In this formula, the Wigner $3j$-symbols can be
 related to the Clebcsh-Gordan coefficients in the usual
 way. \cite{threej}
 For a given angular momentum cutoff $\Lambda$,
 the quantity $\calM_{lm;l'm'}$ can be viewed as the
 matrix element of a linear operator $\hat{M}$ in
 a vector space $\calH_\Lambda$, which is spanned
 by all harmonic polynomials of degree $l\leq\Lambda$.
 The modified zeta function is formally defined by:
 \be
 \label{eq:zeta_def}
 \calZ_{lm}(s,q^2;\eta_1,\eta_2)=
 \sum_{\bn} {\calY_{lm}(\tilde{\bn})
 \over (\tilde{\bn}^2-q^2)^s}\;.
 \ee
 According to this definition,
 the modified zeta function at the right-hand side of Eq.~(\ref{eq:calM-zeta})
 is formally divergent and needs to be analytically continued.
 Following similar discussions as in Ref. \cite{luscher91:finitea},
 one could obtain a finite expression for the modified zeta
 function which is suitable for numerical evaluation.
 Detailed formulae for $\calZ_{lm}(s,q^2;\eta_1,\eta_2)$ at
 $s=1$ and $s=2$ are derived in the appendix.
 From the analytically continued formula,
 it is obvious from the symmetry of $D_4$ or $D_2$
 that, for $l\leq 4$, the only non-vanishing zeta functions
 at $s=1$ are: $\calZ_{00}$, $\calZ_{20}$, $\calZ_{2\pm 2}$,
 $\calZ_{40}$, $\calZ_{4\pm 2}$ and $\calZ_{4\pm 4}$.
 It is also easy to verify that, if $\eta_1=\eta_2=1$, all of the above
 definitions and formulae reduce to the
 those obtained in Ref.~\cite{luscher91:finitea}.

 In the region where the interaction is vanishing,
 the energy eigenstates of the two-particle system can be
 expressed in terms of
 ordinary spherical Bessel functions:
 \be
 \label{eq:eigenfunction}
 \psi_{lm}(r)=b_{lm}\left[
 \alpha_l(k)j_l(kr)+\beta_l(k)n_l(kr)
 \right]\;,
 \ee
 for some constants $b_{lm}$. Also, this energy
 eigenfunction coincide with a singular periodic
 solution of the Helmholtz equation in this region.
 Comparison of Eq.~(\ref{eq:eigenfunction}) with
 Eq.~(\ref{eq:expand_Ylm}) then yields:
 \ba
 \label{eq:compare}
 b_{lm}\alpha_l(k) &=& \sum^\Lambda_{l'=0}\sum^{l'}_{m'=-l'}
 v_{l'm'}{(-)^{l'}k^{l'+1}\over 4\pi}\calM_{l'm';lm}\;,
 \nonumber \\
 b_{lm}\beta_l(k)&=&v_{lm}{(-)^{l}k^{l+1}\over 4\pi}
 \;,
 \ea
 with $l=0,1,\cdots,\Lambda$. Note that this equation
 can be viewed as a linear equation in a vector
 space $\calH_\Lambda$, which is space of all complex
 vectors whose components are $v_{lm}$ with $l=0,1,\cdots,\Lambda$
 and $m=-l,\cdots,l$. Matrix elements $\calM_{l'm';lm}$ can be
 viewed as the matrix element of an operator $M$ in vector
 space $\calH_\Lambda$.
 The scattering phases of the two particles are
 related to the coefficients $\alpha_l(k)$ and
 $\beta_l(k)$ via:
 \be
 e^{2i\delta_l(k)}={\alpha_l(k)+i\beta_l(k)
 \over \alpha_l(k)-i\beta_l(k)}\;.
 \ee
 Therefore, non-trivial solution to Eq.~(\ref{eq:compare}) requires
 that:
 \be
 \label{eq:results_in_H_Lambda}
 \det\left[e^{2i\delta}-U\right]=0\;,\;\;
 U=(M+i)/(M-i)\;.
 \ee
 This gives the general relation between the energy
 eigenvalue of a two-particle eigenstate in a finite
 box with the corresponding scattering phases.

 \section{Symmetry of an asymmetric box}
 \label{sec:symmetry}

 The general result~(\ref{eq:results_in_H_Lambda})
 obtained in the previous section
 can be further simplified when we consider irreducible
 representations of the symmetry group of the box.
 \begin{table}[thb]
 \caption{Basis polynomials in terms of spherical harmonics
 for various irreducible representations $\Gamma$
 of the symmetry group $D_4$ and $D_2$ up to angular momentum $l=4$.
 \label{tab:basis}}
 \begin{center}
 \vspace{2mm}
 \begin{tabular}{|c||c|c||c|c|}
 \hline
     & \multicolumn{2}{c||}{Group $D_4$}
     & \multicolumn{2}{c|}{Group $D_2$} \\    \cline{2-5}
 $l$  & $\Gamma$ & Basis polynomials & $\Gamma$ & Basis polynomials\\
 \hline
 $0$ & $A^+_1$ & $|0\rangle$ & $A^+$  & $|0\rangle =\calY_{00}$ \\
 \hline
 $1$ & $A^-_2$ & $|1\rangle$ & $B^-_1$& $|1\rangle =\calY_{10}$ \\
     & $E^-$   & $(|\tilde{1}\rangle ,|\oneudot{1}{\sim}\rangle)$
     & $B^-_2$; $B^-_3$ & $|\tilde{1}\rangle=(\calY_{11}+\calY_{1-1})$;
                          $|\oneudot{1}{\sim}\rangle=(\calY_{11}-\calY_{1-1})$ \\
 \hline
 $2$ & $A^+_1$ & $|2\rangle$ & $A^+$& $|2\rangle=\calY_{20}$ \\
      & $E^+$   & $(|\tilde{2}\rangle ,|\oneudot{2}{\sim}\rangle)$
     & $B^+_3$; $B^+_2$ & $|\tilde{2}\rangle=(\calY_{21}+\calY_{2-1})$;
                          $|\oneudot{2}{\sim}\rangle=(\calY_{21}-\calY_{2-1})$\\
     & $B^+_1$; $B^+_2$ & $|\bar{2}\rangle$; $|\oneudot{2}{-}\rangle$
     & $A^+$;   $B^+_1$ & $|\bar{2}\rangle=(\calY_{22}+\calY_{2-2})$;
                          $|\oneudot{2}{-}\rangle=(\calY_{22}-\calY_{2-2})$\\
 \hline
  $3$ & $A^-_2$ & $|3\rangle$   & $B^-_1$   & $|3\rangle=\calY_{30}$ \\
      & $E^-$            & $(|\tilde{3}\rangle ,|\oneudot{3}{\sim}\rangle)$
      & $B^-_2$; $B^-_3$ & $|\tilde{3}\rangle=(\calY_{31}+\calY_{3-1})$;
                           $|\oneudot{3}{\sim}\rangle=(\calY_{31}-\calY_{3-1})$ \\
      & $B^-_2$; $B^-_1$ & $|\bar{3}\rangle$; $|\oneudot{3}{-}\rangle$
      & $B^-_1$; $A^-$   & $|\bar{3}\rangle=(\calY_{32}+\calY_{3-2})$;
                           $|\oneudot{3}{-}\rangle=(\calY_{32}-\calY_{3-2})$ \\
      & $E^-$            & $(|\dot{3}\rangle,|\oneudot{3}{\cdot}\rangle)$
      & $B^-_2$; $B^-_3$ & $|\dot{3}\rangle=(\calY_{33}+\calY_{3-3})$;
                           $|\oneudot{3}{\cdot}\rangle=(\calY_{33}-\calY_{3-3})$ \\
 \hline
  $4$ & $A^+_1$ & $|4\rangle$   & $A^+$   & $|4\rangle=\calY_{40}$ \\
      & $E^+$            & $(|\tilde{4}\rangle ,|\oneudot{4}{\sim}\rangle)$
      & $B^+_3$; $B^+_2$ & $|\tilde{4}\rangle=(\calY_{41}+\calY_{4-1})$;
                           $|\oneudot{4}{\sim}\rangle=(\calY_{41}-\calY_{4-1})$ \\
      & $B^+_1$; $B^+_2$ & $|\bar{4}\rangle$; $|\oneudot{4}{-}\rangle$
      & $A^+$;   $B^+_1$ & $|\bar{4}\rangle=(\calY_{42}+\calY_{4-2})$;
                           $|\oneudot{4}{-}\rangle=(\calY_{42}-\calY_{4-2})$ \\
      & $E^+$            & $(|\dot{4}\rangle,|\oneudot{4}{\cdot}\rangle)$
      & $B^+_3$; $B^+_2$ & $|\dot{4}\rangle=(\calY_{43}+\calY_{4-3})$;
                           $|\oneudot{4}{\cdot}\rangle=(\calY_{43}-\calY_{4-3})$ \\
      & $A^+_1$; $A^+_2$ & $|\tcirc{4}\rangle$; $|\oneudot{4}{\circ}\rangle$
      & $A^+$;   $B^+_1$ & $|\tcirc{4}\rangle=(\calY_{44}+\calY_{4-4})$;
                           $|\oneudot{4}{\circ}\rangle=(\calY_{44}-\calY_{4-4})$ \\
      \hline
 \end{tabular}
 \end{center}
 \end{table}
 We know that energy eigenstates in a box can be
 characterized by their transformation properties under
 the symmetry group of the box.
 For this purpose, one has to decompose the representations
 of the rotational group with angular momentum $l$ into
 irreducible representations of the corresponding symmetry
 group of the box. For an asymmetric box, the relevant symmetry group
 is either $D_4$ if $\eta_1=\eta_2\neq 1$, or $D_2$ if
 $\eta_1\neq\eta_2\neq 1$.
 In a given symmetry sector, denoted by its irreducible
 representation $\Gamma$, the representation of the
 rotational group with angular momentum $l$ is decomposed into
 irreducible representations of $D_4$ or $D_2$. This decomposition
 may contain the irreducible representation $\Gamma$.
 We may pick our basis as: $|\Gamma,\alpha;l,n\rangle$.
 Here $\alpha$ runs from $1$ to $dim(\Gamma)$, the dimension
 of the irreducible representation $\Gamma$. Label $n$ runs from
 $1$ to the total number of occurrence of $\Gamma$ in
 the decomposition of rotational group representation with
 angular momentum $l$. The matrix $\hat{M}$ is diagonal with
 respect to $\Gamma$ and $\alpha$ by Schur's lemma.
 For convenience, we have listed these basis polynomials
 in terms of spherical harmonics $\calY_{lm}(\bfr)$ in
 Table~\ref{tab:basis}. The corresponding shorthand notation
 for these basis are also listed.

 For the two-particle eigenstate in the symmetry sector $\Gamma$
 in a box of particular symmetry (either $D_4$ or $D_2$),
 the energy eigenvalue, $E=k^2/(2\mu)$ with $\mu$ being
 the reduced mass of the two particles, is determined by:
 \be
 \label{eq:main_result}
 \det [e^{2i\delta}-\hat{U}(\Gamma)]=0 \;,\;\;
 \hat{U}(\Gamma)=(\hat{M}(\Gamma)+i)/(\hat{M}(\Gamma)-i)\;.
 \ee
 Here $\hat{M}(\Gamma)$ represents a linear operator in
 the vector space $\calH_\Lambda(\Gamma)$.
 This vector space is spanned by all complex vectors
 whose components are $v_{ln}$, with $l\leq\Lambda$,
 and $n$ runs from $1$ to the number of occurrence of
 $\Gamma$ in the decomposition of representation with angular
 momentum $l$, see Ref.~\cite{luscher91:finitea} for details.
 To write out more explicit formulae,
 one therefore has to consider decompositions
 of the rotational group representations under appropriate symmetries.

 We first consider the case $\eta_1=\eta_2$. The basic
 symmetry group is $D_4$, which has $4$ one-dimensional
 (irreducible) representations: $A_1$, $A_2$, $B_1$, $B_2$ and a
 two-dimensional irreducible representation $E$.
 \footnote{The notations of the irreducible representations
 of group $D_4$ and $D_2$ that we adopt here is taken from
 Ref.~\cite{landau:quantum_mechanics_book}, chapter XII.}%
 The representations of the rotational group
 are decomposed according to:
 \ba
 \label{eq:decomposition_D4}
 {\mathbf 0} &=& A^+_1\;,\nonumber \\
 {\mathbf 1} &=& A^-_2+E^-\;,\nonumber \\
 {\mathbf 2} &=& A^+_1+B^+_1+B^+_2+E^+\;, \\
 {\mathbf 3} &=& A^-_2+B^-_1+B^-_2+E^-+E^-\;,\nonumber \\
 {\mathbf 4} &=& A^+_1+A^+_1+A^+_2+B^+_1+B^+_2+E^++E^+\;,\nonumber
 \ea
 In most lattice calculations, the symmetry sector that is
 easiest to investigate is the invariant sector: $A^+_1$.
 We therefore will focus on this particular symmetry sector.
 As is seen,  up to $l\leq 4$, $s$-wave, $d$-wave and $g$ wave contribute to this sector.
 This corresponds to {\em four} linearly independent, homogeneous polynomials
 with degrees not more than $4$, which are
 invariant under $D_4$.
 From Table~\ref{tab:basis} we see that
 the four polynomials can be identified as the basis:
 $|0\rangle$, $|2\rangle$, $|4\rangle$ and
 $|\tcirc{4}\rangle$, respectively.
 \footnote{Our convention for the spherical harmonics are
 taken from Ref.~\cite{jackson:book}.}
 Therefore, we can write out the four-dimensional reduced matrix
 $\calM(A^+_1)$ whose matrix elements are denoted as:
 $\calM(A^+_1)_{ll'}=m_{ll'}=m_{l'l}$, with $l$ and $l'$ takes values
 in $0$, $2$, $4$ and $\tcirc{4}$, respectively.
 Using the general formula~(\ref{eq:calM-zeta}), it is
 straightforward to work out these reduced matrix elements
 in terms of matrix elements $\calM_{lm;l'm'}$. These
 are given explicitly in appendix B.
 We find that, in the case of $D_4$ symmetry,
 Eq.~(\ref{eq:main_result}) becomes:
 \be
 \label{eq:4by4}
 \left|\begin{array}{cccc}
 \cot\delta_0-m_{00} & m_{02} & m_{04} & m_{0\tcirc{4}} \\
 m_{20} & \cot\delta_2-m_{22} & m_{24} & m_{2\tcirc{4}} \\
 m_{40} & m_{42} & \cot\delta_4-m_{44} & m_{4\tcirc{4}} \\
 m_{\tcirc{4}0} & m_{\tcirc{4}2} & m_{\tcirc{4}4} &
 \cot\delta_4-m_{\tcirc{4}\tcirc{4}} \end{array}\right|=0
 \;.
 \ee

 For the case $\eta_1\neq\eta_2$, the symmetry
 group becomes $D_2$ which has only $4$ one-dimensional
 irreducible representations: $A$, $B_1$, $B_2$ and $B_3$.
 The decomposition~(\ref{eq:decomposition_D4}) is replaced by:
 \ba
 \label{eq:decomposition_D2}
 {\mathbf 0}&=&A^+\;,\nonumber \\
 {\mathbf 1}&=&B^-_1+B^-_2+B^-_3\;,\nonumber \\
 {\mathbf 2}&=&A^++A^++B^+_1+B^+_2+B^+_3\;, \\
 {\mathbf 3}&=&A^-+B^-_1+B^-_1+B^-_2+B^-_2+B^-_3+B^-_3\;, \nonumber \\
 {\mathbf 4}&=&A^++A^++A^++B^+_1+B^+_1+B^+_2+B^+_2+B^+_3+B^+_3\;, \nonumber
 \ea
 So, up to $l\leq 4$, $A^+$ occurs {\em six} times:
 once in $l=0$ and twice in $l=2$ and three times in
 $l=4$. The corresponding
 basis polynomials can be taken as:
 $|0\rangle$, $|2\rangle$, $|\bar{2}\rangle$, $|4\rangle$,
 $|\bar{4}\rangle$ and $|\tcirc{4}\rangle$, respectively.
 The reduced matrix $\hat{M}(A^+)$ is six-dimensional
 with matrix elements $\calM_{ll'}=m_{ll'}$.
 The explicit expressions for these matrix elements can by
 found in appendix B. The relation
 between the energy eigenvalue and the scattering phase
 is similar to Eq.~(\ref{eq:4by4}) except that the matrix
 becomes a $6\times 6$ matrix:
 \be
 \label{eq:6by6}
 \left|\begin{array}{cccccc}
 \xi_0-m_{00} & m_{02} & m_{0\bar{2}} & m_{04} & m_{0\bar{4}} & m_{0\tcirc{4}} \\
 m_{20} & \xi_2-m_{22} & m_{2\bar{2}} & m_{24} & m_{2\bar{4}} & m_{2\tcirc{4}} \\
   m_{\bar{2}0} & m_{\bar{2}2} & \xi_2-m_{\bar{2}\bar{2}}
 & m_{\bar{2}4} & m_{\bar{2}\bar{4}} & m_{\bar{2}\tcirc{4}} \\
 m_{40} & m_{42} & m_{4\bar{2}} & \xi_4-m_{44} & m_{4\bar{4}} & m_{4\tcirc{4}} \\
 m_{\bar{4}0} & m_{\bar{4}2} & m_{\bar{4}\bar{2}} &
 m_{\bar{4}4} & \xi_4-m_{\bar{4}\bar{4}} & m_{\bar{4}\tcirc{4}}\\
 m_{\tcirc{4}0} & m_{\tcirc{4}2} & m_{\tcirc{4}\bar{2}} &
 m_{\tcirc{4}4} & m_{\tcirc{4}\bar{4}} & \xi_4-m_{\tcirc{4}\tcirc{4}}
 \end{array}\right|=0
 \;,
 \ee
 where we have used the simplified
 notation: $\xi_l(q)=\cot\delta_l(q)$.
 In principle, if even higher angular momentum are desired,
 similar formulae can be derived.

 \section{Low momentum expansion and the scattering length}
 \label{sec:small_q}

 Usually, scattering phases with higher angular momentum
 are much smaller than phases with lower angular momentum.
 This is particularly true in low-momentum scattering.
 It is well-known that for small relative momentum $k$,
 the scattering phases behave like:
 \be
 \tan\delta_l(k) \sim a_lk^{2l+1} \;,
 \ee
 for small $k$. Therefore, we anticipate that, in the low-momentum limit,
 scattering phases with small $l$ will dominate the scattering
 process. If we treat the $d$-wave and $g$-wave scattering
 phases as small perturbations, we find that for the
 $D_4$ symmetry, Eq.~(\ref{eq:4by4}) can be simplified to:
 \be
 \label{eq:result_D4}
 \cot\delta_0 -m_{00}=
  {m^2_{02}\over \cot\delta_2-m_{22}}
 +{m^2_{04}\over \cot\delta_4-m_{44}}
 +{m^2_{0\bar{4}}\over \cot\delta_4-m_{\bar{4}\bar{4}}}
 \;.
 \ee
 For the symmetry $D_2$, similar to Eq.~(\ref{eq:result_D4}),
 Eq.~(\ref{eq:6by6}) reads:
 \ba
 \label{eq:result_D2}
 \cot\delta_0 -m_{00}&=&
  {m^2_{02}\over \cot\delta_2-m_{22}}
 +{m^2_{0\bar{2}}\over \cot\delta_2-m_{\bar{2}\bar{2}}}
 \nonumber \\
 &+&{m^2_{04}\over \cot\delta_4-m_{44}}
  +{m^2_{0\tilde{4}}\over \cot\delta_4-m_{\tilde{4}\tilde{4}}}
  +{m^2_{0\bar{4}}\over \cot\delta_4-m_{\bar{4}\bar{4}}} \;.
 \ea
 If the $d$-wave and $g$ wave phase shift were
 small enough, it is easy to check that
 both Eq~(\ref{eq:result_D4}) and Eq~(\ref{eq:result_D2})
 simplifies to:
 \be
 \label{eq:simplify}
 \cot\delta_0(k)=m_{00}={\calZ_{00}(1,q^2;\eta_1,\eta_2)
 \over \pi^{3/2}\eta_1\eta_2 q}\;.
 \ee
 For the general case,
 Eq.~(\ref{eq:result_D4}) and Eq.~(\ref{eq:result_D2}) offer
 the desired relation between the energy eigenvalues in
 the $A^+_1$ sector and the scattering phases for the cases
 $\eta_1=\eta_2$ and $\eta_1\neq\eta_2$, respectively.
 It is easy to verify that,
 in both Eq.~(\ref{eq:result_D4}) and  Eq.~(\ref{eq:result_D2}),
 contributions that appear in the right-hand side of the equations
 are smaller by a factor of $q^2$ compared with $m_{00}$ on the left-hand side.
 They are negligible as long as the relative momentum $q$ is small enough.
 Therefore, in both cases, the $s$-wave scattering length $a_0$ will be
 determined by the zero momentum limit of Eq.~(\ref{eq:simplify}).

 \begin{figure}[htb]
 \begin{center}
 \includegraphics[width=\textwidth,angle=0]{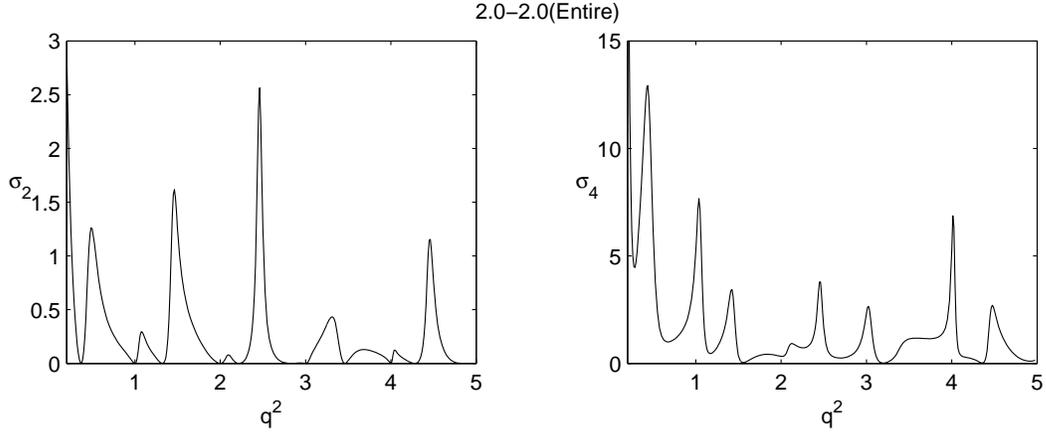}
 \end{center}
 \caption{The functions $\sigma_2(q^2)$ and $\sigma_4(q^2)$
 as a function of $q^2$ are plotted for
 both $D_4$  symmetry with parameters $\eta_1=\eta_2=2$.
 \label{fig:sigma_D4}}
 \end{figure}
 \begin{figure}[htb]
 \begin{center}
 \includegraphics[width=\textwidth,angle=0]{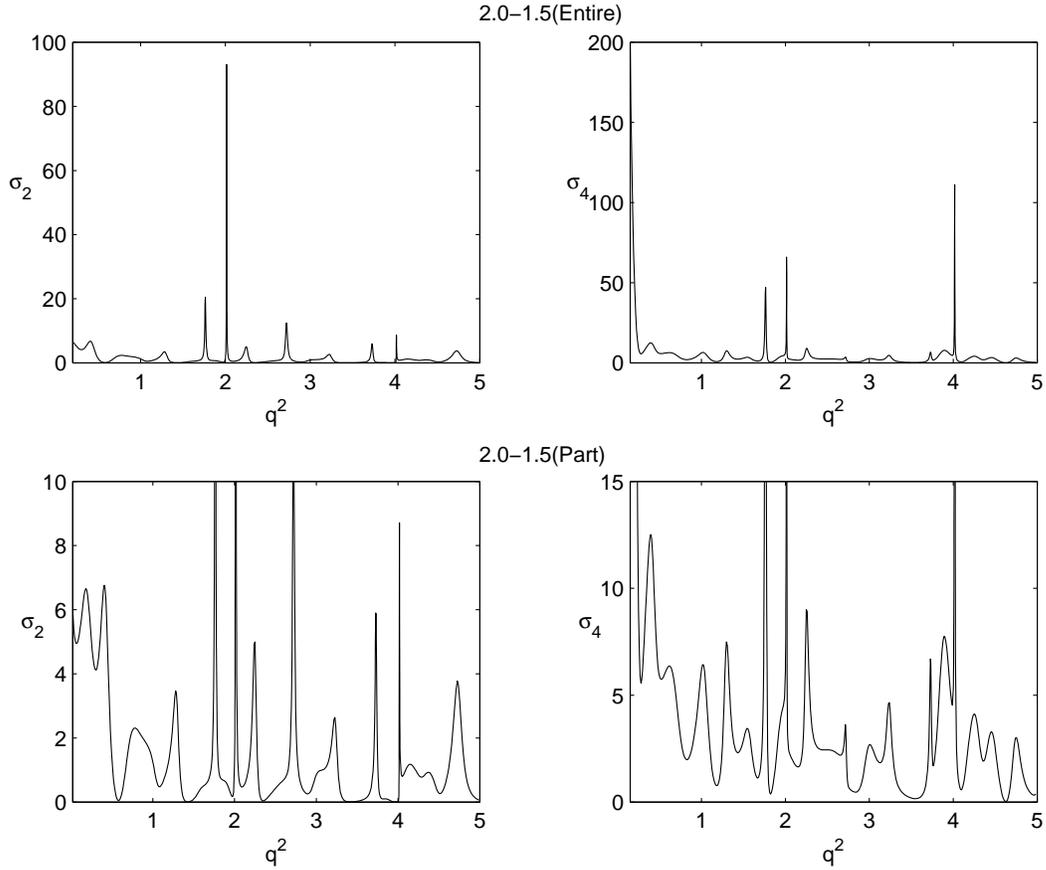}
 \end{center}
 \caption{The functions $\sigma_2(q^2)$ and $\sigma_4(q^2)$
 as a function of $q^2$ are plotted for
 both $D_2$ symmetry with parameters $\eta_1=2$, $\eta_2=1.5$.
 \label{fig:sigma_D2}}
 \end{figure}
 It is also possible to work out the corrections due to higher
 scattering phases to the $s$-wave scattering phase.
 For example, we have:
 \be
 n\pi-\delta_0(q)=\phi(q)
 +\sigma_2(q)\tan\delta_2(q)
 +\sigma_4(q)\tan\delta_4(q)\;,
 \ee
 where the angle $\phi(q)$ is defined to via:
 $-\tan\phi(q)=1/m_{00}(q)$. The functions
 $\sigma_2(q)$ and $\sigma_4(q)$ represents the
 sensitivity of higher scattering phases.
 For $D_4$ and $D_2$ symmetry, they are given by:
 \ba
   \sigma_2(q)&=&\left\{\begin{array}{lc}
   m^2_{02}/(1+m^2_{00})\;, & \mbox{for group $D_4$.} \\
   (m^2_{02}+ m^2_{0\bar{2}})/(1+m^2_{00}) \;, & \mbox{for group $D_2$.}
   \end{array}\right. \\
   \sigma_4(q)&=&\left\{\begin{array}{lc}
   (m^2_{04}+m^2_{0\bar{4}})/( 1+m^2_{00})\;, & \mbox{for group $D_4$.}\\
   (m^2_{04}+ m^2_{0\bar{4}}+m^2_{0\tilde{4}})/(1+m^2_{00})\;, &
   \mbox{for group $D_2$.}
   \end{array}\right.
 \ea
 The functions $\sigma_2(q^2)$ and $\sigma_4(q^2)$ can be
 calculated using the matrix elements given in Appendix B.
 In Fig.~\ref{fig:sigma_D4} and Fig.~\ref{fig:sigma_D2},
 these functions are plotted versus $q^2$ for
 the case of $D_4$ symmetry ($\eta_1=\eta_2=2$) and
 for the case of $D_2$ with $\eta_1=2.0$, $\eta_2=1.5$,
 respectively.
 It is seen that the functions $\sigma_2(q^2)$ and $\sigma_4(q^2)$
 remain finite for all $q^2 >0$. For some particular values
 of $q^2$, however, these functions can become quite large
 in magnitude. This is due to almost coincidence of singularities of
 the numerator and denominator in matrix elements $m_{0i}$
 which happens for some choices of $\eta_1$ and $\eta_2$.
 For values of $q^2$ away from these values, the
 functional values of $\sigma_2$ and $\sigma_4$
 remain moderate. In Fig.~\ref{fig:sigma_D2},
 the lower two panels in the plot are simply
 the same function as in the upper panels with
 the scale of the vertical axis being magnified,
 in order to show the detailed variation of the
 functions.

 We remark here that, in principle,
 the corrections due to scattering
 phases with higher $l$ can be estimated
 from lattice calculations as well. From Table~\ref{tab:basis}
 it is seen that, for lattices with $D_4$ symmetry,
 by inspecting energy eigenstates with
 $E^+$, $B^+_1$ or $B^+_2$ symmetry on the lattice,
 one can get an estimate
 for the $d$-wave scattering phase $\delta_2$ which
 dominates these symmetry sectors.
 Similarly, for lattices with $D_2$ symmetry,
 the eigenstates with $B^+_1$, $B^+_2$ or $B^+_3$ symmetry
 should be studied.
 It is also interesting to note that
 for lattices with $D_4$ symmetry,
 if we study the eigenstates with $A^+_2$ symmetry,
 then the leading contribution comes from $g$-wave
 scattering phase $\delta_4$.

 We now come to the discussion of scattering length.
 For a large box, a large $L$ expansion of the formulae can be deduced.
 Using Eq.~(\ref{eq:simplify}) and following similar
 derivations as in Ref.~\cite{luscher91:finitea},
 we find that the $s$-wave scattering length $a_0$
 is related to the energy difference in a generic
 rectangular box via:
 \be
 \label{eq:scattering_length}
 \delta E=-{2\pi  a_0\over \eta_1\eta_2\mu L^3}
 \left[
 1+c_1(\eta_1,\eta_2)\left({a_0\over L}\right)
  +c_2(\eta_1,\eta_2)\left({a_0\over L}\right)^2
  +\cdots
 \right]\;.
 \ee
 Here, $\mu$ designates the reduced mass of the
 two particles whose mass values are $m_1$ and $m_2$,
 respectively. Energy shift $\delta E\equiv E-m_1-m_2$ where
 $E$ is the energy eigenvalue of the two-particle state.
 Functions $c_1(\eta_1,\eta_2)$ and $c_2(\eta_1,\eta_2)$ are
 given by:
 \ba
 c_1(\eta_1,\eta_2)&=&{\hat{Z}_{00}(1,0;\eta_1,\eta_2)
 \over\pi\eta_1\eta_2} \;,
 \nonumber \\
 c_2(\eta_1,\eta_2)&=&{\hat{Z}^2_{00}(1,0;\eta_1,\eta_2)-
 \hat{Z}_{00}(2,0;\eta_1,\eta_2)\over
 (\pi\eta_1\eta_2)^2}\;,
 \ea
 where the subtracted zeta function is defined as:
 \be
 \label{eq:sub_zeta_def}
 \hat{Z}_{00}(s,q^2;\eta_1,\eta_2)=
 \sum_{|\tilde{\bn}|^2\neq q^2} {1 \over (\tilde{\bn}^2-q^2)^s}\;.
 \ee
 \begin{table}[htb]
 \caption{Numerical values for the subtracted zeta functions
 and the coefficients $c_1(\eta_1,\eta_2)$ and $c_2(\eta_1,\eta_2)$
 under some typical topology. The three-dimensional rectangular
 box has a size $L_1=\eta_1L$, $L_2=\eta_2L$ and $L_3=L$.
 \label{tab:numerical_values}}
 \begin{center}
 \begin{tabular}{c|l|l|r|r|r|r}
 \hline
 $L_1:L_2:L_3$ & $\eta_1$ & $\eta_2$
 & $\hat{Z}_{00}(1,0;\eta_1,\eta_2)$ & $\hat{Z}_{00}(2,0;\eta_1,\eta_2)$%
 & $c_1(\eta_1,\eta_2)$ & $c_2(\eta_1,\eta_2)$ \\
 \hline
 $1:1:1$ & $1$    & $1$   & $-8.913633$  & $16.532316$ & $-2.837297$ & $6.375183$\\
 $6:5:4$ & $1.5$ & $1.25$ & $-12.964476$ & $41.526870$ & $-2.200918$ & $3.647224$\\
 $4:3:2$ & $2$  & $1.5$   & $-16.015122$ & $91.235227$ & $-1.699257$ & $1.860357$\\
 $3:2:2$ & $1.5$    & $1$ & $-10.974332$ & $32.259457$ & $-2.328826$ & $3.970732$\\
 $2:1:1$ & $2$    & $1$   & $-11.346631$ & $63.015304$ & $-1.805872$ & $1.664979$\\
 $3:3:2$ & $1.5$  & $1.5$ & $-14.430365$ & $53.784051$ & $-2.041479$ & $3.091200$\\
 $2:2:1$ & $2$  & $2$     & $-18.430516$ & $137.771800$ & $-1.466654$ & $1.278623$\\
 \hline
 \end{tabular}
 \end{center}
 \end{table}
 In Table~\ref{tab:numerical_values}, we have listed
 numerical values for the coefficients
 $c_1(\eta_1,\eta_2)$ and $c_2(\eta_1,\eta_2)$
 under some typical topology. In the first column of
 the table, we tabulated the ratio for the three sides
 of the box: $\eta_1:\eta_2:1$. Note that for $\eta_1=\eta_2=1$,
 these two functions reduce to the old numerical values
 for the cubic box which had been used in earlier
 scattering length calculations.

 It can be shown that contaminations from higher angular
 momentum scattering phases come in at even higher
 powers of $1/L$. For low relative momenta,
 the $d$-wave scattering phase behaves
 like: $\tan\delta_2(q)\sim a_2k^5=a_2(2\pi /L)^5q^5$, with
 $a_2$ being the $d$-wave scattering length.
 If we treat the effects due to $\tan\delta_2$
 perturbatively, we see from Eq.~(\ref{eq:result_D4}) and
 Eq.~(\ref{eq:result_D2}) that,
 in Eq.~(\ref{eq:scattering_length}), functions $c_1$ and $c_2$
 receive contributions that are proportional to $(a_2/L^5)$, which
 is of higher order in $1/L$ for large $L$.

 \section{Conclusions}
 \label{sec:conclude}

 In this paper, we have studied two-particle scattering states
 in a generic rectangular box with periodic boundary conditions.
 The relations of the energy eigenvalues and the scattering phases
 in the continuum are found. These formulae
 can be viewed as a generalization of
 the well-known L\"uscher's formulae.
 In particular, we show that the $s$-wave scattering length is
 related to the energy shift by a simple formula, which is a
 direct generalization of the corresponding formula in the case of
 cubic box. We argued that this asymmetric topology
 might be useful in practice since it provides more available low-lying
 momentum modes in a finite box, which will be
 advantageous in the study of scattering phase shifts at
 non-zero three momenta in hadron-hadron scattering and
 possibly also in other applications.

 \appendix
 \section{Appendix A}
 \label{app:zeta}

 In this appendix, some explicit formulae for the
 modified zeta function defined in Eq.~(\ref{eq:zeta_def}) will
 be given. For convenience of analytic continuation, one
 first defines the heat kernel:
 \be
 \label{eq:heat_kernel}
 \calK(t,\bx)={1\over (2\pi)^3}\sum_\bn e^{i\tn\cdot\bx-t\tn^2}
 = {\eta_1\eta_2\over (4\pi t)^{3/2}}
 \sum_\bn e^{-{1\over 4t}\left(\bx-2\pi\hn\right)^2}\;.
 \ee
 Here the first equality is the definition of the heat kernel
 $\calK(t,\bx)$ while the second follows from Poisson's
 identity. The summations which appear in this formula
 runs over all three-dimensional integers $\bn\in Z^3$ and
 the notations $\tn$ and $\hn$ are defined as in
 Eq.~(\ref{eq:bn_hn_def}).
 It is evident that the heat kernel $\calK(t,\bx)$ is
 periodic in $\bx$ with period $2\pi\hn$, i.e.
 $\calK(t,\bx+2\pi\hn)=\calK(t,\bx)$.
 Given a positive number $\Lambda>0$, we define the
 truncated heat kernel $\calK^\Lambda(t,\bx)$ as:
 \be
 \label{eq:K_Lambda}
 \calK^\Lambda(t,\bx)=\calK(t,\bx)-
 {1\over (2\pi)^3}\sum_{|\tn|\leq\Lambda}
 e^{i\tn\cdot\bx-t\tn^2}\;.
 \ee
 We may apply the operator $\calY_{lm}(-i\nabla_\bx)$ to
 the heat kernels defined above. These are denoted as:
 \be
 \calK_{lm}(t,\bx)=\calY_{lm}(-i\nabla_\bx)\calK(t,\bx)\;,
 \;\;
 \calK^\Lambda_{lm}(t,\bx)=\calY_{lm}(-i\nabla_\bx)\calK^\Lambda(t,\bx)\;,
 \ee

 Using heat kernels defined above, one could rewrite the
 modified zeta function as:
 \be
 \calZ_{lm}(s,q^2;\eta_1,\eta_2)=\sum_{|\tn|\leq\Lambda}
 {\calY_{lm}(\tn) \over (\tn^2-q^2)^s}
 +{(2\pi)^3\over\Gamma(s)}\int^\infty_0 dt t^{s-1}
 e^{tq^2}\calK^\Lambda_{lm}(t,\bzero) \;.
 \ee
 This expression is convergent as long as $Re(s) >(l+3)/2$.
 Close to $t=0$, the function
 $e^{tq^2}\calK^\Lambda_{lm}(t,\bzero)$ behaves like:
 \be
 e^{tq^2}\calK^\Lambda_{lm}(t,\bzero)\sim
 {\delta_{l0}\delta_{m0}\eta_1\eta_2\over (4\pi)^2 t^{3/2}}
 +\calO (t^{-1/2})\;.
 \ee
 We may therefore analytically continue the zeta
 function by:
 \ba
 \label{eq:zeta_continued}
 \!\!\!\!\!\!\!\!\!
 \calZ_{lm}(s,q^2;\eta_1,\eta_2)&=&\sum_{|\tn|\leq\Lambda}
 {\calY_{lm}(\tn) \over (\tn^2-q^2)^s}
 +{(2\pi)^3\over\Gamma(s)}\left\{
 {\delta_{l0}\delta_{m0}\eta_1\eta_2\over (4\pi)^2 (s-3/2)}\right.
 \nonumber \\
 &+&\int^1_0 dt t^{s-1}\left[
 e^{tq^2}\calK^\Lambda_{lm}(t,\bzero)
 -{\delta_{l0}\delta_{m0}\eta_1\eta_2\over (4\pi)^2 t^{3/2}}
 \right] \nonumber \\
 &+&\left.\int^\infty_1 dt t^{s-1}
 e^{tq^2}\calK^\Lambda_{lm}(t,\bzero)
 \right\} \;,
 \ea
 which is a valid expression as long as $Re(s) >1/2$.
 This completes the process of analytic continuation for
 the modified zeta functions. Using
 the explicit expression for $\calK^\Lambda_{lm}(t,\bzero)$,
 the integral from $0$ to $1$ in the above formula can be
 further simplified. After some manipulations, we find that
 the results, when combined with the finite summation, is
 convergent for all $\Lambda$. Therefore, we can now
 send $\Lambda$ to infinity and drop the third integral.
 Since we are only concerned with modified zeta functions
 at $s=1$ or $s=2$, we will only give the expressions
 for these two cases. The final expression may be written as:
 \ba
 \label{eq:zeta_final}
 \calZ_{lm}(s,q^2;\eta_1,\eta_2)&=&e^{q^2}\sum_\bn
 \left[{\calY_{lm}(\tn) \over (\tn^2-q^2)^s}
 +{(s-1)\calY_{lm}(\tn)\over (\tn^2-q^2)^{s-1}}\right]
 e^{-\tn^2}
 \nonumber \\
 &+& {\pi\eta_1\eta_2\over(2s-3)}\delta_{l0}\delta_{m0}
 +{\pi\eta_1\eta_2\over 2}\delta_{l0}\delta_{m0}
 \int^1_0 dt t^{s-5/2}(e^{tq^2}-1)
 \nonumber \\
 &+& \pi\eta_1\eta_2\int^1_0 dt t^{s-5/2}\left[
 \sum_{\bn\neq\bzero}\calY_{lm}\left(-i{\pi\over t}\hn\right)
 e^{tq^2}e^{-{\pi^2\over t}\hn^2}
 \right]\;.
 \ea
 Note that this expression is valid only for $s=1$ or $s=2$.

 \section{Appendix B}
 \label{app:matrix elements}

 \begin{table}[htb]
 \caption{Reduced matrix elements $\calM_{ll'}(\Gamma)$
 for the symmetry group $D_4$ for irreducible
 representations $A^+_1$, $A^-_2$ and $E^-$.
 Here we have adopted the basis defined
 in Table~\ref{tab:basis}. For all representations,
 only the non-vanishing matrix elements up to
 $l=4$ are listed.
 \label{tab:matrix_D4A}}
 \begin{center}
 \begin{tabular}{|c|c|c|c|c}
 \hline
 $\Gamma$ & $|l\rangle$ & $|l'\rangle$ & $\calM_{ll'}(\Gamma)$ \\
 \hline
 $A^+_1$ & $0$  & $0$   & $\calW_{00}$ \\
         & $2$  & $0$   & $-\calW_{20}$ \\
         & $2$  & $2$   & $\calW_{00}+{2\sqrt{5}\over 7}\calW_{20}+{6\over 7}\calW_{40}$ \\
         & $4$  & $0$   & $\calW_{40}$ \\
         & $4$  & $2$   & $-{6\over 7}\calW_{20}-{20\sqrt{5}\over 77}\calW_{40}%
                           -{15\sqrt{65}\over 143}\calW_{60}$ \\
         & $4$  & $4$   & $\calW_{00}+{20\sqrt{5}\over 77}\calW_{20}%
                           +{486\over 1001}\calW_{40}+{20\sqrt{13}\over 143}\calW_{60}%
                           +{490\sqrt{17}\over 2431}\calW_{80}$ \\
         & $\tcirc{4}$ & $0$ & ${1\over\sqrt{2}}(\calW_{44}+\calW_{4-4})$ \\
         & $\tcirc{4}$ & $2$ & ${2\sqrt{10}\over 11}(\calW_{44}+\calW_{4-4})%
                           -{15\over 11\sqrt{26}}(\calW_{64}+\calW_{6-4})$ \\
         & $\tcirc{4}$ & $4$ & ${27\sqrt{2}\over 143}(\calW_{44}+\calW_{4-4})%
                           -{6\sqrt{10}\over 11\sqrt{13}}(\calW_{64}+\calW_{6-4})%
                           +{21\sqrt{1870}\over 4862}(\calW_{84}+\calW_{8-4})$ \\
         & $\tcirc{4}$ & $\tcirc{4}$ & $\calW_{00}-{4\sqrt{5}\over 11}\calW_{20}%
                           +{54\over 143}\calW_{40}-{4\over 11\sqrt{13}}\calW_{60}%
                           +{7\over 143\sqrt{17}}\calW_{80}%
                           +{21\sqrt{5}\over \sqrt{4862}}(\calW_{88}+\calW_{8-8})$ \\
 \hline
 $A^-_2$ & $1$  & $1$   & $\calW_{00}+{2\over \sqrt{5}}\calW_{20}$ \\
         & $3$  & $1$   & $-{3\sqrt{3}\over\sqrt{35}}\calW_{20}%
                           -{4\over \sqrt{21}}\calW_{40}$ \\
         & $3$  & $3$   & $\calW_{00}+{4\over 3\sqrt{5}}\calW_{20}%
                           +{6\over 11}\calW_{40}+{100\over 33\sqrt{13}}\calW_{60}$ \\
 \hline
 $E^-$ & $\tilde{\oneudot{1}{\sim}}$ & $\tilde{\oneudot{1}{\sim}}$
       & $\calW_{00}-{1\over \sqrt{5}}\calW_{20}%
        \mp{\sqrt{3}\over\sqrt{10}}(\calW_{22}+\calW_{2-2})$ \\
       & $\tilde{\oneudot{3}{\sim}}$ & $\tilde{\oneudot{3}{\sim}}$
       & $\calW_{00}+{1\over \sqrt{5}}\calW_{20}%
       +{1\over 11}\calW_{40}-{25\over 11\sqrt{13}}\calW_{60}$\\
       & &
       & $\mp{\sqrt{2}\over\sqrt{15}}(\calW_{22}+\calW_{2-2})%
       \mp{\sqrt{10}\over 11}(\calW_{42}+\calW_{4-2})%
       \mp{5\sqrt{35}\over 11\sqrt{39}}(\calW_{62}+\calW_{6-2})$  \\
       & $\tilde{\oneudot{3}{\sim}}$ & $\tilde{\oneudot{1}{\sim}}$
       & $-{3\sqrt{2}\over \sqrt{35}}\calW_{20}%
       +{\sqrt{2}\over \sqrt{7}}\calW_{40}%
       \mp{\sqrt{3}\over 2\sqrt{35}}(\calW_{22}+\calW_{2-2})%
       \pm{\sqrt{5}\over 2\sqrt{7}}(\calW_{42}+\calW_{4-2})$\\
       & $\dot{\oneudot{3}{\cdot}}$ & $\dot{\oneudot{3}{\cdot}}$
       & $\calW_{00}-{\sqrt{5}\over 3}\calW_{20}%
       +{3\over 11}\calW_{40}-{5\over 33\sqrt{13}}\calW_{60}%
       \mp{35\over \sqrt{3003}}(\calW_{66}+\calW_{6-6})$\\
       & $\dot{\oneudot{3}{\cdot}}$ & $\tilde{\oneudot{1}{\sim}}$
       & $-{3\over 2\sqrt{7}}(\calW_{22}+\calW_{2-2})%
       +{1\over 2\sqrt{21}}(\calW_{42}+\calW_{4-2})
       \pm{1\over\sqrt{3}}(\calW_{44}+\calW_{4-4})$\\
       & $\dot{\oneudot{3}{\cdot}}$ & $\tilde{\oneudot{3}{\sim}}$
       & $-{\sqrt{2}\over 6}(\calW_{22}+\calW_{2-2})%
       +{3\sqrt{6}\over 22}(\calW_{42}+\calW_{4-2})
       -{5\sqrt{7}\over 33\sqrt{13}}(\calW_{62}+\calW_{6-2})$\\
       & &
       & $\pm{\sqrt{42}\over 22}(\calW_{44}+\calW_{4-4})
       \mp{5\sqrt{35}\over 11\sqrt{78}}(\calW_{64}+\calW_{6-4})$\\
 \hline
 \end{tabular}
 \end{center}
 \end{table}

 \newpage
 \begin{table}[htb]
 \caption{Reduced matrix elements $\calM_{ll'}(\Gamma)$
 for the symmetry group $D_4$ for irreducible
 representations $E^+$, $B^\pm_1$, $B^\pm_2$, and $A^+_2$.
 \label{tab:matrix_D4B}}
 \begin{center}
 \begin{tabular}{|c|c|c|c|c}
 \hline
 $\Gamma$ & $|l\rangle$ & $|l'\rangle$ & $\calM_{ll'}(\Gamma)$ \\
 \hline
 $E^+$ & $\tilde{\oneudot{2}{\sim}}$ & $\tilde{\oneudot{2}{\sim}}$
       & $\calW_{00}+{\sqrt{5}\over 7}\calW_{20}-{4\over 7}\calW_{40}%
                                     \mp{\sqrt{30}\over 14}(\calW_{22}+\calW_{2-2})%
                                     \mp{\sqrt{10}\over 7}(\calW_{42}+\calW_{4-2})$  \\
       & $\tilde{\oneudot{4}{\sim}}$ & $\tilde{\oneudot{2}{\sim}}$
       & $-{\sqrt{30}\over 7}\calW_{20}-{5\sqrt{2}\over 77}\calW_{40}
                                       +{10\sqrt{6}\over 11\sqrt{13}}\calW_{60}$ \\
       & &
       & $\mp{\sqrt{5}\over 14}(\calW_{22}+\calW_{2-2})%
          \pm{9\sqrt{15}\over 154}(\calW_{42}+\calW_{4-2})%
          \pm{2\sqrt{70}\over 11\sqrt{13}}(\calW_{62}+\calW_{6-2})$\\
       & $\tilde{\oneudot{4}{\sim}}$ & $\tilde{\oneudot{4}{\sim}}$
       & $\calW_{00}+{17\sqrt{5}\over 77}\calW_{20}
         +{243\over 1001}\calW_{40}-{1\over 11\sqrt{13}}\calW_{60}
         -{392\over 143\sqrt{17}}\calW_{80}$ \\
       & &
       & $\mp{5\sqrt{30}\over 77}(\calW_{22}+\calW_{2-2})%
         \mp{81\sqrt{10}\over 1001}(\calW_{42}+\calW_{4-2})$\\
       & &
       & $\mp{\sqrt{105}\over 11\sqrt{13}}(\calW_{62}+\calW_{6-2})%
         \mp{21\sqrt{140}\over 143\sqrt{17}}(\calW_{82}+\calW_{8-2})$\\
       & $\dot{\oneudot{4}{\cdot}}$ & $\tilde{\oneudot{2}{\sim}}$
       & $-{\sqrt{5}\over 2\sqrt{7}}(\calW_{22}+\calW_{2-2})%
          -{5\sqrt{15}\over 22\sqrt{7}}(\calW_{42}+\calW_{4-2})%
          +{2\sqrt{10}\over 11\sqrt{13}}(\calW_{62}+\calW_{6-2})$\\
      & &
      & $\pm{\sqrt{15}\over 11}(\calW_{44}+\calW_{4-4})
         \pm{10\sqrt{3}\over 11\sqrt{13}}(\calW_{64}+\calW_{6-4})$\\
       & $\dot{\oneudot{4}{\cdot}}$ & $\dot{\oneudot{4}{\cdot}}$
       & $\calW_{00}-{\sqrt{5}\over 11}\calW_{20}
        -{81\over 143}\calW_{40}+{17\over 11\sqrt{13}}\calW_{60}
        -{56\over 143\sqrt{17}}\calW_{80}$\\
      & &
      & $\mp{\sqrt{21}\over\sqrt{143}}(\calW_{66}+\calW_{6-6})
         \mp{14\sqrt{3}\over \sqrt{2431}}(\calW_{86}+\calW_{8-6})$\\
       & $\dot{\oneudot{4}{\cdot}}$ & $\tilde{\oneudot{4}{\sim}}$
       & $-{3\sqrt{15}\over 11\sqrt{14}}(\calW_{22}+\calW_{2-2})
          +{27\over 143\sqrt{14}}(\calW_{42}+\calW_{4-2})$\\
      & &
      & $+{3\sqrt{15}\over 11\sqrt{13}}(\calW_{62}+\calW_{6-2})
         -{42\sqrt{5}\over 143\sqrt{17}}(\calW_{82}+\calW_{8-2})$\\
      & &
      & $\pm{27\sqrt{10}\over 286}(\calW_{44}+\calW_{4-4})
         \pm{3\over 11\sqrt{26}}(\calW_{64}+\calW_{6-4})
         \mp{42\sqrt{2}\over 13\sqrt{187}}(\calW_{84}+\calW_{8-4})$\\
 \hline
 $B^+_{1/2}$ & $\bar{\oneudot{2}{-}}$  & $\bar{\oneudot{2}{-}}$
       & $\calW_{00}-{2\sqrt{5}\over 7}\calW_{20}+{1\over 7}\calW_{40}%
       \pm{\sqrt{5}\over\sqrt{14}}(\calW_{44}+\calW_{4-4})$ \\
      & $\bar{\oneudot{4}{-}}$  & $\bar{\oneudot{4}{-}}$
      & $\calW_{00}+{8\sqrt{5}\over 77}\calW_{20}-{27\over 91}\calW_{40}%
       -{2\over \sqrt{13}}\calW_{60}+{196\over 143\sqrt{17}}\calW_{80}$ \\
      & &
      & $\pm{81\sqrt{5}\over 143\sqrt{14}}(\calW_{44}+\calW_{4-4})
        \pm {3\sqrt{14}\over 11\sqrt{13}}(\calW_{64}+\calW_{6-4})
        \pm {21\sqrt{14}\over 13\sqrt{187}}(\calW_{84}+\calW_{8-4})$ \\
      & $\bar{\oneudot{4}{-}}$  & $\bar{\oneudot{2}{-}}$
      & $-{\sqrt{15}\over 7}\calW_{20}+{30\sqrt{3}\over 77}\calW_{40}%
       -{5\sqrt{3}\over 11\sqrt{13}}\calW_{60}$ \\
      & &
      & $\pm{\sqrt{30}\over 11\sqrt{7}}(\calW_{44}+\calW_{4-4})
         \mp{5\sqrt{42}\over 22\sqrt{13}}(\calW_{64}+\calW_{6-4})$ \\
 \hline
 $B^-_{2/1}$ & $\bar{\oneudot{3}{-}}$  & $\bar{\oneudot{3}{-}}$
       & $\calW_{00}-{7\over 11}\calW_{40}+{10\over 11\sqrt{13}}\calW_{60}%
       \pm{\sqrt{70}\over 22}(\calW_{44}+\calW_{4-4})
       \pm{5\sqrt{14}\over 11\sqrt{13}}(\calW_{64}+\calW_{6-4})$ \\
 \hline
 $A^+_2$ & $\oneudot{4}{\circ}$ & $\oneudot{4}{\circ}$
 & $\calW_{00}-{4\sqrt{5}\over 11}\calW_{20}%
 +{54\over 143}\calW_{40}-{4\over 11\sqrt{13}}\calW_{60}%
 +{7\over 143\sqrt{17}}\calW_{80}%
 -{21\sqrt{5}\over \sqrt{4862}}(\calW_{88}+\calW_{8-8})$ \\
 \hline
 \end{tabular}
 \end{center}
 \end{table}

 In this appendix, we list the reduced matrix elements
 that appeared in the formulae in the main text.
 We define:
 \be
 \label{eq:W_def}
 \calW_{lm}(1,q^2;\eta_1,\eta_2)\equiv
 {\calZ_{lm}(1,q^2;\eta_1,\eta_2)
 \over \pi^{3/2}\eta_1\eta_2q^{l+1}}\;.
 \ee
 Using this notation, we now proceed to list relevant
 reduced matrix elements $\calM_{ln;l'n'}(\Gamma)$ for
 a given symmetry sector $\Gamma$.
 For the case of $D_4$ symmetry, up to angular momentum $l=4$,
 the non-vanishing matrix
 elements $\calM_{ln;l'n'}(\Gamma)$ are listed
 in Table~\ref{tab:matrix_D4A}, Table~\ref{tab:matrix_D4B}.

 For the case of $D_2$ symmetry, by comparing
 with Table~\ref{tab:basis}, one easily sees
 that the corresponding matrix elements are in fact the
 same as listed in Table~\ref{tab:matrix_D4A} and
 Table~\ref{tab:matrix_D4B}. The only difference is that
 the name of the irreducible representations are changed.
 For example, the matrix elements listed under
 $E^-$ for $D_4$ symmetry are in fact corresponding
 matrix elements for the representation $B^-_2$ and $B^-_3$
 for the $D_2$ symmetry, as suggested by Table~\ref{tab:basis}.


\begin{thebibliography}{10}

\bibitem{luscher86:finiteb}
M.~L{\"u}scher.
\newblock {\em Commun. Math. Phys.}, 105:153, 1986.

\bibitem{luscher90:finite}
M.~L{\"u}scher and U.~Wolff.
\newblock {\em Nucl. Phys. B}, 339:222, 1990.

\bibitem{luscher91:finitea}
M.~L{\"u}scher.
\newblock {\em Nucl. Phys. B}, 354:531, 1991.

\bibitem{luscher91:finiteb}
M.~L{\"u}scher.
\newblock {\em Nucl. Phys. B}, 364:237, 1991.

\bibitem{Zimmermann94}
M.~Goeckeler, H.A. Kastrup, J.~Westphalen, and F.~Zimmermann.
\newblock {\em Nucl. Phys. B}, 425:413, 1994.

\bibitem{gupta93:scat}
R.~Gupta, A.~Patel, and S.~Sharpe.
\newblock {\em Phys. Rev. D}, 48:388, 1993.

\bibitem{fukugita95:scat}
M.~Fukugita, Y.~Kuramashi, H.~Mino, M.~Okawa, and A.~Ukawa.
\newblock {\em Phys. Rev. D}, 52:3003, 1995.

\bibitem{jlqcd99:scat}
S.~Aoki et~al.
\newblock {\em Nucl. Phys. (Proc. Suppl.) B}, 83:241, 2000.

\bibitem{JLQCD02:pipi_length}
JLQCD Collaboration.
\newblock {\em Phys. Rev. D}, 66:077501, 2002.

\bibitem{chuan02:pipiI2}
C.~Liu, J.~Zhang, Y.~Chen, and J.P. Ma.
\newblock {\em Nucl. Phys. B}, 624:360, 2002.

\bibitem{juge03:pipi_length}
K.J. Juge.
\newblock {\em hep-lat/0309075}, 2003.

\bibitem{CPPACS03:pipi_phase}
CP-PACS Collaboration.
\newblock {\em Phys. Rev. D}, 67:014502, 2003.

\bibitem{ishizuka03:pipi_length}
N.~Ishizuka and T.~Yamazaki.
\newblock {\em hep-lat/0309168}, 2003.

\bibitem{CPPACS03:pipi_phase_unquench}
CP-PACS Collaboration.
\newblock {\em hep-lat/0309155}, 2003.

\bibitem{ishizuka02:Kpipi_review}
N.~Ishizuka.
\newblock {\em hep-lat/0209108}, 2002.

\bibitem{threej}
M.~Rotenberg, R.~Bivins, N.~Metropolis, and J.K.~Wooten Jr.
\newblock {\em The 3-j and 6-j Symbols}.
\newblock The Technology Press, Massachusetts Institute of Technology,
  Cambridge, Massachusetts, USA, 1959.

\bibitem{landau:quantum_mechanics_book}
L.D. Landau and E.M. Lifshitz.
\newblock {\em Quantum Mechanics, 3rd ed.}
\newblock Pergamon Press, Oxford, UK.

\bibitem{jackson:book}
J.~D. Jackson.
\newblock {\em Classical Electrodynamics}.
\newblock John Wiley \& Sons Inc., New York, USA, 1975.

\end{thebibliography}

\end{document}